\documentclass[12pt,aps,prb,preprint]{revtex4}

\usepackage{graphicx}
\usepackage{dcolumn}

\begin{document}

\title{Accurate calculation of eigenvalues and eigenfunctions. I: Symmetric
potentials}
\author{Francisco M. Fern\'andez}

\affiliation{INIFTA (UNLP, CCT La Plata-CONICET), Divisi\'{o}n Qu\'{i}mica
Te\'{o}rica, Diag. 113 y 64 (S/N), Sucursal 4, Casilla de Correo 16, 1900 La
Plata, Argentina}

\begin{abstract}
We describe a method for the calculation of accurate energy eigenvalues and
expectation values of observables of separable quantum--mechanical models.
We discuss the application of the approach to one--dimensional anharmonic
oscillators with symmetric potential--energy functions.
\end{abstract}

\maketitle

\section{Introduction}

Some time ago we developed a new method for the calculation of eigenvalues
and eigenfunctions of the Schr\"{o}dinger equation for separable problems.
We first applied it to bound states\cite
{FMT89a,FMT89b,F92,FG93,FGZ93,F95b,FT96} but later we found that it also
applied to resonances or metastable states\cite{F95,F95c,F96,F96b,F08}, as
well as to some strange complex eigenvalues\cite{F08b}.

Although the method only applies to separable quantum--mechanical problems
it exhibits many interesting features that may be suitable for the
discussion at a graduate, or advance undergraduate level. The teaching of
the approach may be greatly facilitated by available computer algebra
systems (CAS) that enable one to carry out the necessary calculation without
much effort or special programming ability.

In this paper we introduce the method and apply it to the one--dimensional
Schr\"{o}dinger equation with a symmetric potential. In order to make this
article more comprehensive, in Sec.~\ref{sec:HO} we discuss the standard
power--series treatment of the harmonic oscillator. In Sec.~\ref{sec:RPMHO}
we develop the method from the known solutions of the harmonic oscillator.
In Sec.~\ref{sec:AHO} we generalize it and apply it to some anharmonic
oscillators with symmetric potentials. In Sec.~\ref{sec:mult_root} we
discuss the phenomenon of multiplicity of roots that is one of the
particular features of the approach. In Sec.~\ref{sec:spu_root} we address
the occurrence of meaningful spurious roots which is another peculiar
feature of the method. In Sec.~\ref{sec:eigenfunct} we consider the
eigenfunctions and the calculation of expectation values. In Sec.~we apply
the method to a simple double--well oscillator. In Sec.~\ref{sec:conclusions}
we draw some conclusions on the teaching and utility of the procedure.

\section{The harmonic oscillator\label{sec:HO}}

In order to facilitate the introduction of the new approach and the
understanding of most of its special features we first discuss an standard
approach to the harmonic oscillator. For simplicity we assume that we have
rewritten the Hamiltonian operator in dimensionless form as $\hat{H}=\hat{p}%
^{2}+V(\hat{x})$ that in the coordinate representation reads $\hat{H}%
=-d^{2}/dx^{2}+V(x)$. Thus, the Schr\"{o}dinger equation for the harmonic
oscillator reads
\begin{equation}
\psi ^{\prime \prime }(x)+(E-x^{2})\psi (x)=0,\;\psi (x\rightarrow \pm
\infty )=0  \label{eq:Schro_HO}
\end{equation}
The logarithmic derivative $y(x)=\psi ^{\prime }(x)/\psi (x)$ satisfies the
Riccati equation $y^{\prime }(x)+y(x)^{2}-x^{2}+E=0$. For large values of $%
|x|$ we find that $y(x)\sim \pm x$ and $\psi (x)\sim e^{\pm x^{2}/2}$. We
appreciate that there are two solutions that we may call convergent $\psi
_{c}(x)$ and divergent $\psi _{d}(x)$ with asymptotic behaviours $\psi
_{c}(x)\sim e^{-x^{2}/2}$ and $\psi _{d}(x)\sim e^{x^{2}/2}$ at large values
of $|x|$. The most general solution is a linear combination of both: $\psi
(x)=C_{1}(E)\psi _{c}(x)+C_{2}(E)\psi _{d}(x)$. The allowed values of the
energy $E$ are the roots $E_{0},E_{1},\ldots $ of $C_{2}(E)=0$. Such
eigenvalues $E_{n}$ are consistent with a square integrable function $%
\int_{-\infty }^{\infty }|\psi |^{2}dx<\infty $ given in this case by $\psi
_{c}(x)$.

A textbook method for the solution of the Schr\"{o}dinger equation for the
harmonic oscillator is based on a power--series expansion of the solution
that we write as\cite{P68}
\begin{equation}
\psi (x)=e^{-x^{2}/2}\sum_{j=0}c_{j}x^{2j+s}  \label{eq:HO_series}
\end{equation}
where $s$ determines the behaviour of $\psi (x)$ at origin. The coefficients
$c_{j}$ satisfy the recurrence relation $%
(2j+s+1)(2j+s+2)c_{j+1}+[E-1-2(2j+s)]c_{j}=0$, and $c_{0}\neq 0$ provided
that $s=0$ or $s=1$ that give rise to the even or odd states, respectively.

One can prove that for arbitrary values of $E$ the series in equation (\ref
{eq:HO_series}) behaves asymptotically as $e^{x^{2}}$ and the divergent
contribution to the eigenfunction $\psi (x)$ would dominate at large $|x|$.
The eigenfunction will not be square integrable unless the otherwise
infinite series in equation (\ref{eq:HO_series}) becomes a polynomial, which
occurs when $E=E_{n}=2n+1$, $n=2\nu +s=0,1,\ldots $, $\nu =0,1,\ldots $.
When $E=E_{n}$ then $c_{\nu +1}=0$ and $c_{j}=0$ for all $j\geq \nu +1$. The
eigenfunctions then read
\begin{eqnarray}
\psi _{n}(x) &=&N_{n}x^{s}P_{n}(x)e^{-x^{2}/2},\;P_{n}=\sum_{j=0}^{\nu
}c_{j,n}x^{2j+s}  \nonumber \\
c_{j+1,n} &=&\frac{4(j-\nu )c_{j,n}}{(2j+s+1)(2j+s+2)},\;j=0,1,\ldots ,\nu
\label{eq:HO_Psi_n}
\end{eqnarray}
The polynomials $x^{s}P_{n}(x)$ are proportional to the well known Hermite
ones\cite{P68}, and the explicit form of the normalization factor $N_{n}$ is
unnecessary for our aims.

The brief discussion given above has the sole purpose of calling the
reader's attention on the close relationship between the correct asymptotic
behaviour of the solution of the Schr\"{o}dinger equation at $|x|\rightarrow
\infty $ and the allowed values of the energy, as well as on the fact that $%
s/x-\psi _{n}^{\prime }(x)/\psi _{n}(x)$ is a rational function, analytic at
origin, and can be expanded in a Taylor series:
\begin{equation}
\frac{s}{x}-\frac{\psi _{n}^{\prime }(x)}{\psi _{n}(x)}=\frac{%
xP_{n}(x)-P_{n}^{\prime }(x)}{P_{n}(x)}=\sum_{j=0}f_{j}x^{2j+1}
\label{eq:HO_log_der_1}
\end{equation}

\section{The Riccati--Pad\'{e} method for the Harmonic oscillator \label%
{sec:RPMHO}}

Suppose that we do not know the allowed values of the energy of the harmonic
oscillator and consider the regularized logarithmic derivative of the
eigenfunction
\begin{equation}
f(x)=\frac{s}{x}-\frac{\psi ^{\prime }(x)}{\psi (x)}  \label{eq:f(x)}
\end{equation}
that satisfies the Riccati equation
\begin{equation}
f^{\prime }(x)+\frac{2sf(x)}{x}-f(x)^{2}+x^{2}-E-\frac{s(s-1)}{x^{2}}=0
\label{eq:HO_Riccati}
\end{equation}
where the last term vanishes because $s=0$ or $s=1$ as discussed earlier. In
order to derive this expression we simply differentiate $f(x)$ with respect
to $x$ and use equations (\ref{eq:Schro_HO}) and (\ref{eq:f(x)}) to get rid
of $\psi ^{\prime \prime }(x)/\psi (x)$ and $\psi ^{\prime }(x)/\psi (x)$.
According to what was said above we know that we can expand $f(x)$ in a
Taylor series about the origin
\begin{equation}
f(x)=x\sum_{j=0}^{\infty }f_{j}(E)z^{j},\;z=x^{2}  \label{eq:f(x)_series}
\end{equation}
Its substitution into the Riccati equation (\ref{eq:HO_Riccati}) leads to a
recurrence relation for the coefficients $f_{j}$ (see below) that enables us
to obtain as many of them as necessary. The first few coefficients of this
expansion when $s=0$ are
\begin{eqnarray}
f_{0} &=&E  \nonumber \\
f_{1} &=&\frac{E^{2}-1}{3}  \nonumber \\
f_{2} &=&\frac{2(E^{2}-1)E}{15}  \nonumber \\
f_{3} &=&\frac{(E^{2}-1)(17E^{2}-5)}{315}  \label{eq:HO_f_j}
\end{eqnarray}
Their general form appears to be $f_{j}=(E^{2}-1)Q_{j}(E)$ for $j>0$, where $%
Q_{j}(E)$ is a polynomial function of the energy.

Taking into account that the regularized logarithmic derivative of the
eigenfunction is an exact rational function for the harmonic oscillator when
$E=E_{n}$, as shown by equation (\ref{eq:HO_log_der_1}), we look for a
solution of the form $f(x)=x[M/N](x^{2})$, where
\begin{equation}
\lbrack M/N](z)=\frac{\sum_{j=0}^{M}a_{j}z^{j}}{\sum_{j=0}^{N}b_{j}z^{j}}%
=\sum_{j=0}^{M+N+1}f_{j}(E)z^{j}+O(z^{M+N+2})  \label{eq:[M/N](z)}
\end{equation}
Since we can arbitrarily choose $b_{0}=1$ we are left with $M+N+1$
coefficients of the rational function and the unknown energy as
independently adjustable parameters; therefore we can try to obtain $M+N+2$
exact coefficients of the Taylor series for $f(x)$ as explicitly indicated
in equation (\ref{eq:[M/N](z)}). When $M\geq N$ we easily derive the
following equations:
\begin{eqnarray}
\sum_{k=0}^{\min (j,N)}b_{k}f_{j-k} &=&a_{j},\;j=0,1,\ldots ,M  \nonumber \\
\sum_{k=0}^{N}b_{k}f_{j-k} &=&0,\;j=M+1,M+2,\ldots ,M+N+1  \label{eq:aj_bj}
\end{eqnarray}
We can view the second set as a system of $N+1$ homogeneous equations with $%
N+1$ unknowns $b_{N},b_{N-1},\ldots ,b_{0}$. Therefore, there will be a
nontrivial solution if $E$ is a root of
\begin{equation}
H_{D}^{d}(E)=\left|
\begin{array}{cccc}
f_{M-N+1} & f_{M-N+2} & \cdots & f_{M+1} \\
f_{M-N+2} & f_{M-N+3} & \cdots & f_{M+2} \\
\vdots & \vdots & \ddots & \vdots \\
f_{M+1} & f_{M+2} & \cdots & f_{M+N+1}
\end{array}
\right| =0  \label{eq:Hankel}
\end{equation}
where $d=M-N=0,1,\ldots $ and $D=N+1=2,3,\ldots $ is the dimension of the
Hankel determinant $H_{D}^{d}(E)$. The first Hankel determinants for $s=0$
are:
\begin{eqnarray}
H_{2}^{0}(E) &=&\frac{(E^{2}-1)^{2}(E^{2}-25)}{4725}  \nonumber \\
H_{2}^{1}(E) &=&\frac{(E^{2}-1)^{2}(E^{2}-25)(E^{2}+3)}{297675}  \nonumber \\
H_{3}^{0}(E) &=&\frac{(E^{2}-1)^{3}(E^{2}-25)^{2}(E^{2}-81)}{46414974375}
\nonumber \\
H_{3}^{1}(E) &=&\frac{4(E^{2}-1)^{3}(E^{2}-25)^{2}(E^{2}-81)E}{%
896041080309375}  \label{eq:HO_H_D^d}
\end{eqnarray}
The general form appears to be $H_{D}^{d}=(E^{2}-1)^{D}(E^{2}-25)^{D-1}%
\ldots [E^{2}-(4D-3)^{2}]G_{D}^{d}(E)$ where $G_{D}^{d}(E)$ is a polynomial
function of $E$. We appreciate that

\begin{itemize}
\item  As $D$ increases more exact eigenvalues appear as roots of $%
H_{D}^{d}(E)=0$

\item  The multiplicity of each root increases with $D$

\item  In addition to the ``physical'' roots $E_{n}=2n+1$ there are spurious
ones; for example: $-\left( 2n+1\right) $ (later on we will discuss the
occurrence of such negative roots)

\item  We have obtained the allowed energies without taking into account the
asymptotic behaviour of the eigenfunctions at large $|x|$ explicitly
\end{itemize}

\section{Anharmonic oscillators\label{sec:AHO}}

In this section we generalize the main results derived above for the
harmonic oscillator and develop the method for the Schr\"{o}dinger equation
\begin{equation}
\psi ^{\prime \prime }(x)+[E-V(x)]\psi (x)=0  \label{eq:AHO_Schro}
\end{equation}
where $V(x)$ is analytic at origin and symmetric about this point: $%
V(-x)=V(x)$. Without loss of generality we assume that $V(0)=0$. Since we
can expand $V(x)$ in a Taylor series
\begin{equation}
V(x)=\sum_{j=1}^{\infty }V_{j}x^{2j}  \label{eq:V_series}
\end{equation}
then we can apply the method exactly as indicated above for the harmonic
oscillator. The regularized logarithmic derivative of the eigenfunction $%
f(x)=s/x-\psi ^{\prime }(x)/\psi (x)$ satisfies the Riccati equation
\begin{equation}
f^{\prime }(x)+\frac{2sf(x)}{x}-f(x)^{2}+V(x)-E=0  \label{eq:AHO_Riccati}
\end{equation}
If we expand $f(x)$ in odd--power series as in equation (\ref{eq:f(x)_series}%
) we easily calculate the coefficients $f_{j}$ by means of the recurrence
relation
\begin{eqnarray}
f_{n} &=&\frac{1}{2n+2s+1}\left( \sum_{j=0}^{n-1}f_{j}f_{n-j-1}+E\delta
_{n0}-V_{n}\right) ,\;n=1,2,\ldots  \nonumber \\
f_{0} &=&\frac{E}{2s+1}  \label{eq:fn}
\end{eqnarray}
and then construct the rational approximation (\ref{eq:[M/N](z)}) that leads
to the Hankel determinant (\ref{eq:Hankel}). We do not expect to obtain
exact eigenvalues for the general anharmonic oscillator as we have already
done for the harmonic one, but at least we do expect that the roots of the
Hankel determinants will enable us to estimate the eigenvalues of the
Schr\"{o}dinger equation (\ref{eq:AHO_Schro}). We call this approach
Riccati--Pad\'{e} method (RPM) because it is based on the Riccati equation
for the modified logarithmic derivative of the eigenfunction, and a rational
approximation or Pad\'{e} approximant (\ref{eq:[M/N](z)})\cite{BO78}.

The simplest symmetric anharmonic potential is
\begin{equation}
V(x)=x^{2}+\lambda x^{4}  \label{eq:V_x2_x4}
\end{equation}
This model has been widely studied because the perturbation expansion
\begin{equation}
E(\lambda )=\sum_{j=0}^{\infty }E^{(j)}\lambda ^{j}  \label{eq:E_series}
\end{equation}
is divergent\cite{BO78,BW69,BW71,S70}. A straightforward argument based on
scaling the coordinate of the Schr\"{o}dinger equation shows that the energy
can also be expanded as
\begin{equation}
E(\lambda )=\lambda ^{1/3}\sum_{j=0}^{\infty }e^{(j)}\lambda ^{-2j/3}
\label{eq:x2_x4_asymp_series}
\end{equation}
with a nonzero convergence radius\cite{BO78,S70}.

\begin{figure}[H]
\begin{center}
\includegraphics[width=9cm]{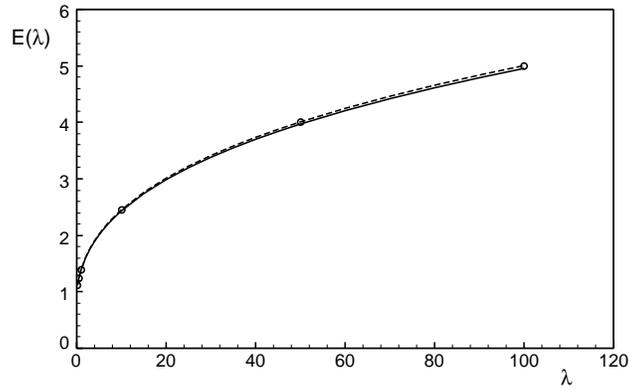}
\end{center}
\caption{Ground--state energy of the anharmonic oscillator (\ref{eq:V_x2_x4}%
). The solid and dashed curves are the analytical lower (\ref
{eq:x2_x4_H(2,0)}) and upper (\ref{eq:x2_x4_H(2,1)}) bounds, respectively,
and the circles mark more accurate results}
\label{fig:anal}
\end{figure}

From $H_{2}^{0}(E)$ we obtain (for concreteness we restrict to even states $%
s=0$)
\begin{equation}
E^{6}-27E^{4}+162E^{3}\lambda +51E^{2}-162E\lambda -189\lambda ^{2}-25=0
\label{eq:x2_x4_H(2,0)}
\end{equation}
Fig.~\ref{fig:anal} shows that this simple expression yields reasonable
results for the ground state of the anharmonic oscillator (\ref{eq:V_x2_x4})
for all $\lambda $ values. The roots $E(\lambda )$ of (\ref{eq:x2_x4_H(2,0)}%
) can be expanded in $\lambda $--power series (\ref{eq:E_series}) and also
in $\lambda ^{-2/3}$--power series (\ref{eq:x2_x4_asymp_series}). To prove
the latter statement, simply substitute $\lambda ^{1/3}W$ for $E$ in
equation (\ref{eq:x2_x4_H(2,0)}) and notice that the roots of the resulting
equation
\begin{equation}
W^{6}-27W^{4}\lambda ^{-2/3}+162W^{3}+51W^{2}\lambda ^{-4/3}-162W\lambda
^{-2/3}-25\lambda ^{-2}-189=0  \label{eq:x2_x4_H(2,0)_W}
\end{equation}
are functions of $\lambda ^{-2/3}$. Besides, when $\lambda \rightarrow
\infty $ we obtain
\begin{equation}
W^{6}+162W^{3}-189=0  \label{eq:x4_W}
\end{equation}
One of its roots $W\approx 1.05$ is quite close to the exact leading
coefficient $e^{(0)}$ (calculated below) and explains why the simple
expression (\ref{eq:x2_x4_H(2,0)}) yields satisfactory results for all
values of $\lambda $.

From $H_{2}^{1}(E)$ we obtain
\begin{equation}
E^{8}-24E^{6}-558E^{5}\lambda -30E^{4}+1404E^{3}\lambda +(666\lambda
^{2}+128)E^{2}-846E\lambda -882\lambda ^{2}-75=0  \label{eq:x2_x4_H(2,1)}
\end{equation}
It has been proved that roots of $H_{D}^{0}(E)$ and $H_{D}^{1}(E)$ are lower
and upper bounds, respectively, to the eigenvalues of the Schr\"{o}dinger
equation with the potential (\ref{eq:V_x2_x4})\cite{FMT89a,FMT89b}. Fig.~\ref
{fig:anal} also shows the approximate ground--state eigenvalue given by Eq. (%
\ref{eq:x2_x4_H(2,1)}) and we appreciate that the upper and lower bounds are
quite close for all values of $\lambda $.

\begin{figure}[H]
\begin{center}
\includegraphics[width=9cm]{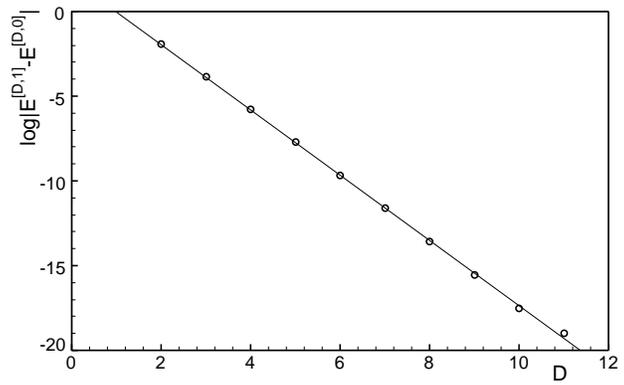}
\end{center}
\caption{Convergence of the upper and lower bounds to the ground--state
energy of the anharmonic oscillator (\ref{eq:V_x^4})}
\label{fig:logUBLB_0}
\end{figure}

In order to test the rate of convergence of the RPM we consider the most
difficult case $\lambda \rightarrow \infty $, sometimes called the
strong--coupling limit. Since $\lim_{\lambda \rightarrow \infty }\lambda
^{-1/3}E(\lambda )=e^{(0)}$ is an eigenvalue of $\hat{H}=\hat{p}^{2}+\hat{x}%
^{4}$ we consider the potential
\begin{equation}
V(x)=x^{4}  \label{eq:V_x^4}
\end{equation}
in what follows. Let $E_{n}^{[D,d]}$ be a sequence of roots of $H_{D}^{d}(E)$
that converges towards the energy eigenvalue $E_{n}$ of the anharmonic
oscillator with the potential (\ref{eq:V_x^4}). As said above, $%
E_{n}^{[D,0]} $ and $E_{n}^{[D,1]}$ converge towards $E_{n}$ from below and
above, respectively, as $D$ increases, giving progressively tighter lower
and upper bounds\cite{FMT89a,FMT89b}. Fig.~\ref{fig:logUBLB_0} shows $\log
\left| E_{0}^{[D,1]}-E_{0}^{[D,0]}\right| $ that is a measure of the rate of
convergence of the bounds towards the ground--state eigenvalue as $D$
increases. From straightforward linear regression we estimate that $\left|
E_{0}^{[D,1]}-E_{0}^{[D,0]}\right| \approx 77.31e^{-4.43D}$, $D=2,3,\ldots $%
. Not many approaches exhibit such remarkable exponential rate of
convergence. We find that $E_{0}^{[11,0]}=1.0603620904841828996$ is exact up
to the last digit and that the accuracy of the RPM results increases with $D$
according to the exponential rate of convergence just given.

The number of zeros of $\psi _{n}(x)$ increases with $n$ and for that reason
we need determinants of greater dimension to take into account the
increasing oscillation of the excited states (think of the rational
approximation to $f(x)$). For the first excited even state $n=2$ we obtain $%
\left| E_{2}^{[D,1]}-E_{2}^{[D,0]}\right| \approx \left( 8.02\times
10^{4}\right) e^{-4.50D}$, $D=3,4,\ldots $ and $%
E_{2}^{[12,0]}=7.4556979379867383922$ is exact up to the last digit.

\section{Multiplicity of roots\label{sec:mult_root}}

\begin{figure}[H]
\begin{center}
\includegraphics[width=9cm]{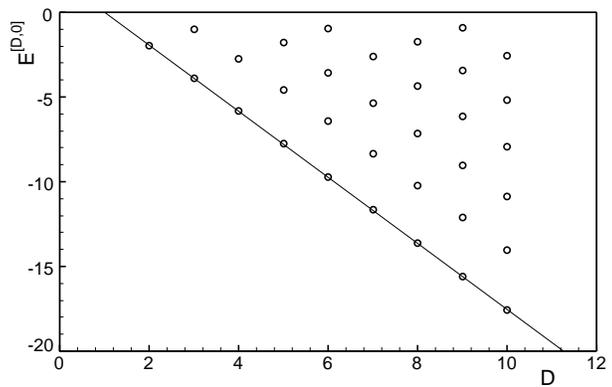}
\end{center}
\caption{Sequences of roots converging towards the ground--state energy of
the anharmonic oscillator (\ref{eq:V_x^4}) as $\log \left|
E_{0}^{[D,0]}-E_{0}^{[\infty,0]}\right| $}
\label{fig:sequences}
\end{figure}

In the case of the harmonic oscillator we saw that the multiplicity of each
root increases with the determinant dimension. The counterpart of this
unusual feature of the RPM in the case of problems that are not exactly
solvable is the occurrence of many roots in the neighbourhood of the chosen
eigenvalue. Fig.~\ref{fig:sequences} shows $\log \left|
E_{0}^{[D,0]}-E_{0}^{[\infty ,0]}\right| $ for the anharmonic oscillator
with the potential (\ref{eq:V_x^4}), where we have arbitrarily chosen $%
E_{0}^{[\infty ,0]}=E_{0}^{[11,0]}$ for this numerical experiment. We
appreciate that there are several parallel sequences of roots converging
towards the ground--state eigenvalue. A straight line $\log \left|
E_{0}^{[D,0]}-E_{0}^{[\infty ,0]}\right| \approx 1.95(1-D)$ marks the
optimal sequence that was chosen above as a lower bound. The strategy for
identifying the optimal sequence from the other ones consists simply in
selecting the root of $H_{D}^{d}(E)=0$ closest to the one chosen previously
from $H_{D-1}^{d}(E)=0$.

For the first even excited state we find $\log \left|
E_{2}^{[D,0]}-E_{2}^{[\infty ,0]}\right| \approx 4.85-1.95D$ which is
parallel to the ground--state one.

\section{Spurious roots\label{sec:spu_root}}

The RPM may also yield roots that are not related to the eigenvalues of the
chosen model. To understand this interesting feature of the approach
consider the change of variables $x=\gamma q$ that transforms $\hat{H}%
=-d^{2}/dx^{2}+V(x)$ into $\hat{H}=\gamma ^{-2}\left[ -d^{2}/dq^{2}+\gamma
^{2}V(\gamma q)\right] $. Therefore, some roots of the Hankel determinant
for $\hat{H}$ will also give us an approximation to $\gamma ^{-2}$ times the
eigenvalues of $-d^{2}/dq^{2}+\gamma ^{2}V(\gamma q)$ provided that $\gamma
^{2}V(\gamma q)$ is a ``reasonable'' potential--energy function\cite{F96}.
For example, in the case of the harmonic oscillator $\gamma ^{2}V(\gamma
q)=\gamma ^{4}q^{2}$ and the choice $\gamma ^{2}=-1$ explains the occurrence
of the negative roots $-\left( 2n+1\right) $ discussed above.

It is interesting to consider the modified Posch--Teller potential\cite{F99}
\begin{equation}
V(x)=-\frac{\lambda (\lambda -1)}{\cosh (x)^{2}}  \label{eq:V_mod_PT}
\end{equation}
with exact eigenvalues $E_{n}(MPT)=-(\lambda -1-n)^{2}$. The change of
coordinates discussed above with $\gamma =i$ yields $-1$ times the
Hamiltonian operator with the Posch--Teller potential\cite{F99}
\begin{equation}
V(q)=\frac{\lambda (\lambda -1)}{\cos (q)^{2}}  \label{eq:V_PT}
\end{equation}
and eigenvalues $E_{n}(PT)=(\lambda +2n)^{2}$. We thus expect sequences of
roots of the Hankel determinant converging towards $E_{n}(MPT)$ and $%
-E_{n}(PT)$\cite{F96}.

Table~\ref{tab:spurious} shows them for $n=0$ and $d=0$. This example is
greatly revealing because the asymptotic behaviour of the bound--state
eigenfunctions of the first model (\ref{eq:V_mod_PT}) is $\psi (x\rightarrow
\pm \infty )=0$, whereas for the second one (\ref{eq:V_PT}) the Dirichlet
boundary conditions $\psi (\pm \pi /2)=0$ are determined by the poles of the
potential--energy function (\ref{eq:V_PT}) at $\pm \pi /2$. We appreciate
that the RPM yields solutions to two models with quite different boundary
conditions simultaneously. This is partly a consequence of the fact that the
RPM does not take the boundary conditions explicitly into account as
discussed earlier in Sec.~\ref{sec:RPMHO}.

\section{Eigenfunctions\label{sec:eigenfunct}}

\begin{figure}[H]
\begin{center}
\includegraphics[width=9cm]{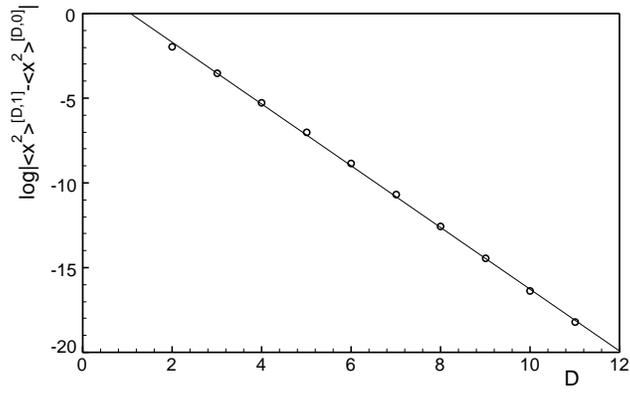}
\end{center}
\caption{Convergence of the upper and lower bounds to $\left\langle
x^{2}\right\rangle $ for the ground state of the anharmonic oscillator with
potential (\ref{eq:V_x^4})}
\label{fig:exval}
\end{figure}

Once we have a sufficiently accurate eigenvalue we obtain the coefficients $%
f_{j}$ and the rational approximation (\ref{eq:[M/N](z)}) to $f(x)$. The
approximate eigenfunction given by
\begin{equation}
\psi (x)=Nx^{s}e^{-\int^{x}f(y)dy}  \label{eq:Psi(x)}
\end{equation}
will be accurate in a neighbourhood of the origin but in general will not
satisfy the boundary conditions at infinity. In principle, one can use the
accurate energy and coefficients to derive an improved expression for the
eigenfunction that behaves satisfactorily at infinity. However, if one needs
to calculate the expectation value of an operator $\hat{A}$ it is preferable
to apply the RPM to $\hat{H}+\beta \hat{A}$ and calculate the slope $%
dE/d\beta $ at $\beta =0$. According to the Hellmann--Feynman theorem we have%
\cite{F00}
\begin{equation}
\frac{dE}{d\beta }(\beta )=\left\langle \hat{A}\right\rangle (\beta )
\label{eq:<A>}
\end{equation}
Straigtforward differentiation of $H_{D}^{d}(E,\beta )=0$ with respect of $%
\beta $ leads to $dH_{D}^{d}/d\beta =(\partial H_{D}^{d}/\partial \beta
)_{E}+(\partial H_{D}^{d}/\partial E)_{\beta }(dE/d\beta )=0$ from which it
follows that
\begin{equation}
\frac{dE}{d\beta }=-\frac{(\partial H_{D}^{d}/\partial \beta )_{E}}{%
(\partial H_{D}^{d}/\partial E)_{\beta }}  \label{eq:de/dbeta_RPM}
\end{equation}
Fig.~\ref{fig:exval} shows the convergence of the RPM values for $%
\left\langle x^{2}\right\rangle $ for the ground state of the anharmonic
oscillator with potential (\ref{eq:V_x^4}) estimated by equations (\ref
{eq:<A>}) and (\ref{eq:de/dbeta_RPM}) at $\beta =0$. In this case we simply
apply the RPM to the anharmonic oscillator with potential--energy function $%
V(x)=x^{4}+\beta x^{2}$ and obtain $\left\langle x^{2}\right\rangle
=0.3620226487886768452$. It is worth noticing that $\left\langle
x^{2}\right\rangle =e^{(1)}$ is the second--leading coefficient of the
series (\ref{eq:x2_x4_asymp_series}).

\section{Double--well\label{sec:DW} oscillator}

The RPM developed above is also suitable for the calculation of the
eigenvalues of the Schr\"{o}dinger equation with symmetric double--well
potentials. For concreteness and simplicity we consider the anharmonic
oscillator with the potential--energy function
\begin{equation}
V(x)=x^{4}+\beta x^{2}  \label{eq:Vx4x2}
\end{equation}
that exhibits two wells when $\beta <0$. They are located at $x_{w}=\pm
\sqrt{-\beta /2}$ and their depth is $V(x_{w})=-\beta ^{2}/4$.

\begin{figure}[H]
\begin{center}
\includegraphics[width=9cm]{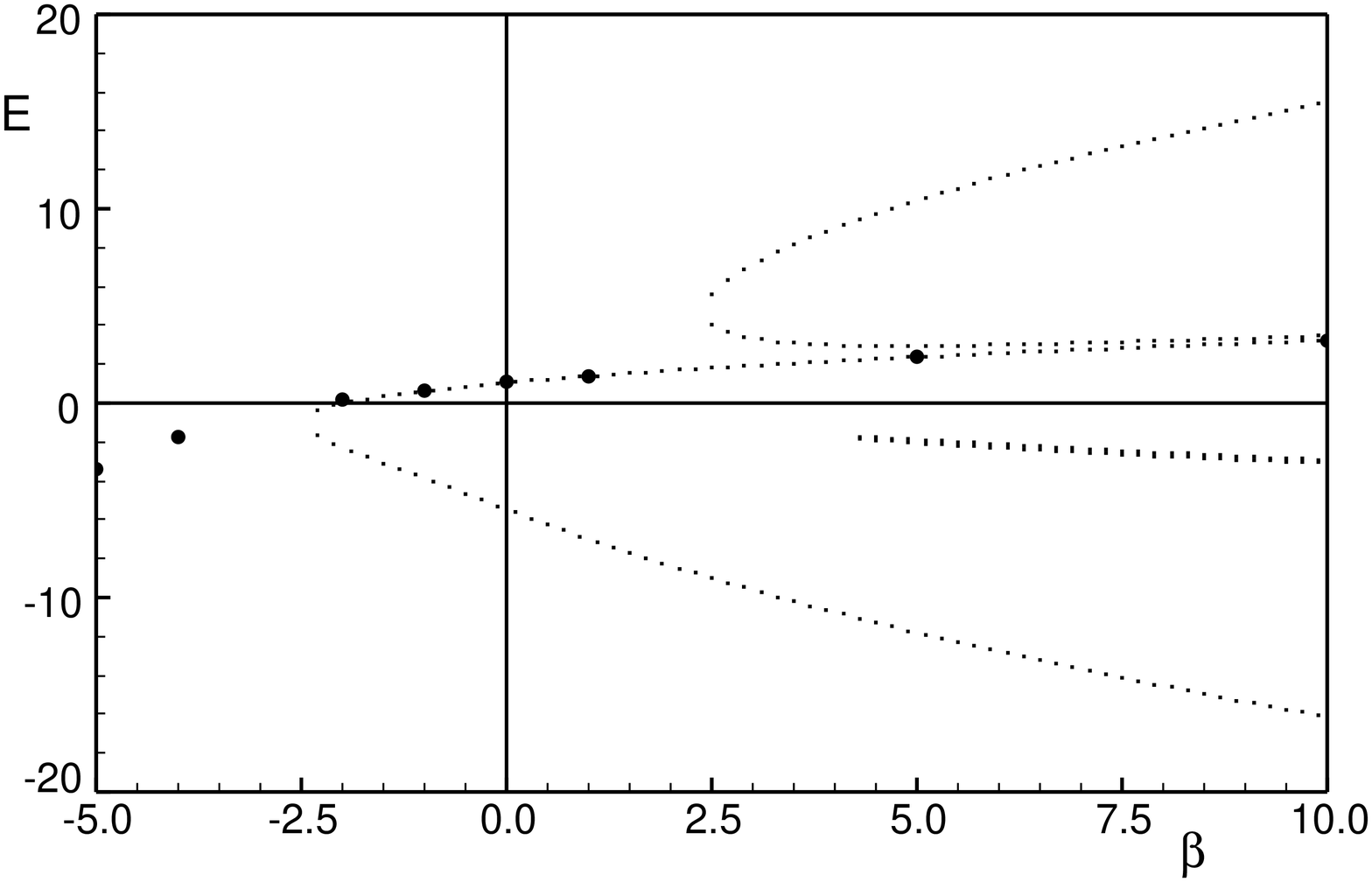}
\end{center}
\caption{Roots $E(\beta)$ of $H_2^0(E,\beta)=$ (points) and accurate results
(filled circles) for the ground state of the double--well oscillator (\ref
{eq:Vx4x2})}
\label{fig:DWH20}
\end{figure}

\begin{figure}[H]
\begin{center}
\includegraphics[width=9cm]{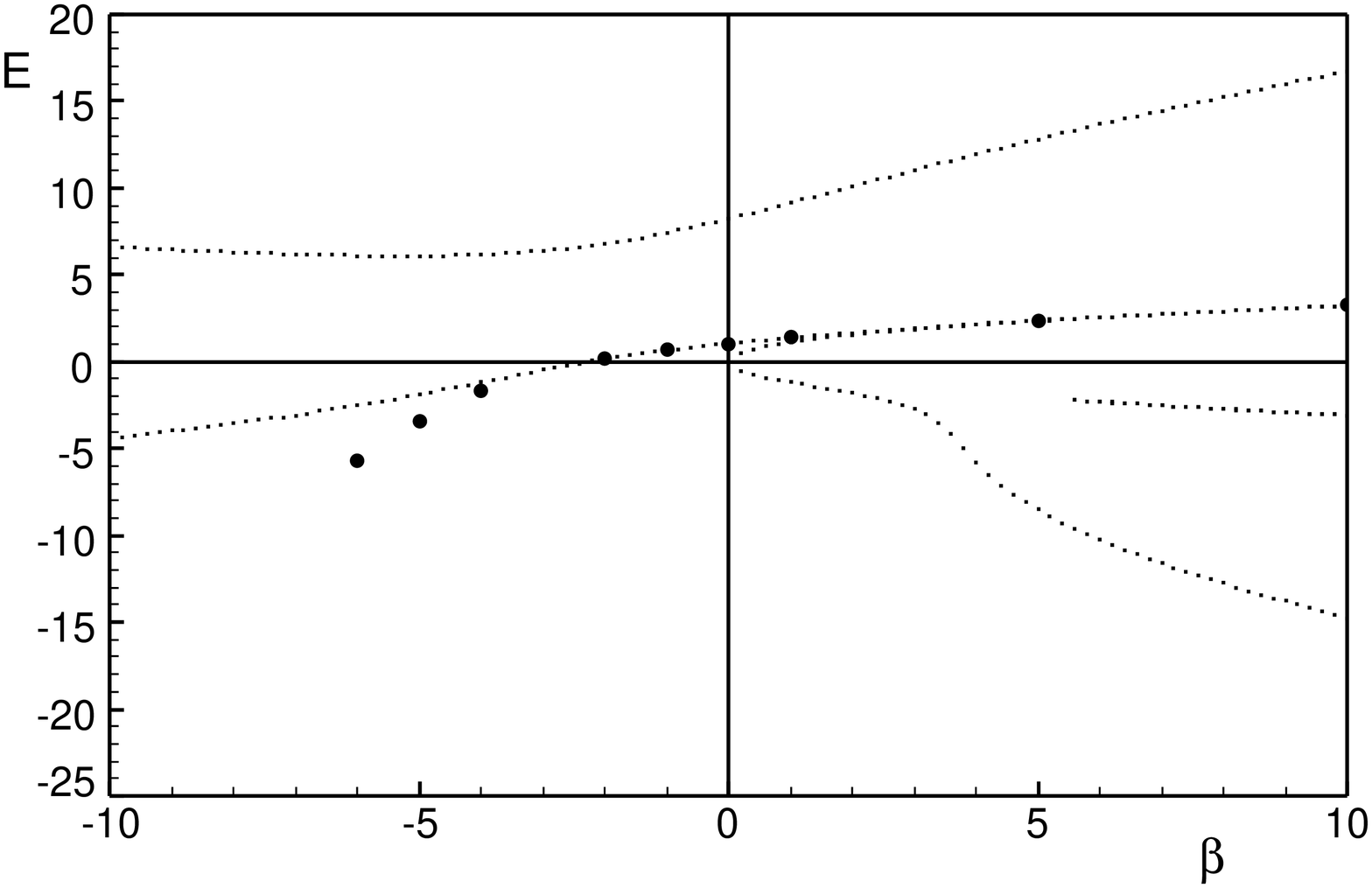}
\end{center}
\caption{Roots $E(\beta)$ of $H_2^1(E,\beta)=$ (points) and accurate results
(filled circles) for the ground state of the double--well oscillator (\ref
{eq:Vx4x2})}
\label{fig:DWH21}
\end{figure}

As in the case of the anharmonic oscillator discussed above we find that the
RPM provides lower and upper bounds: $E_{n}^{[D,0]}<E_{n}<E_{n}^{[D,1]}$. We
first consider the implicit analytic expressions given by $H_{2}^{0}(E,\beta
)=0$, and $H_{2}^{1}(E,\beta )=0$:
\begin{eqnarray}
E^{6}-27E^{4}\beta +162E^{3}+51E^{2}\beta ^{2}-162E\beta -25\beta ^{3}-189
&=&0  \nonumber \\
E^{8}-24E^{6}\beta -558E^{5}-30E^{4}\beta ^{2}+1404E^{3}\beta &&  \nonumber
\\
+(666+128\beta ^{3})E^{2}-846E\beta ^{2}-882\beta -75\beta ^{4} &=&0
\label{eq:DWH2d}
\end{eqnarray}
respectively. Fig.~\ref{fig:DWH20} and \ref{fig:DWH21} shows the roots of
these equations as functions of $\beta $. We appreciate that one branch of
each equation agrees with the accurate results also plotted in those figures.

\begin{figure}[H]
\begin{center}
\includegraphics[width=9cm]{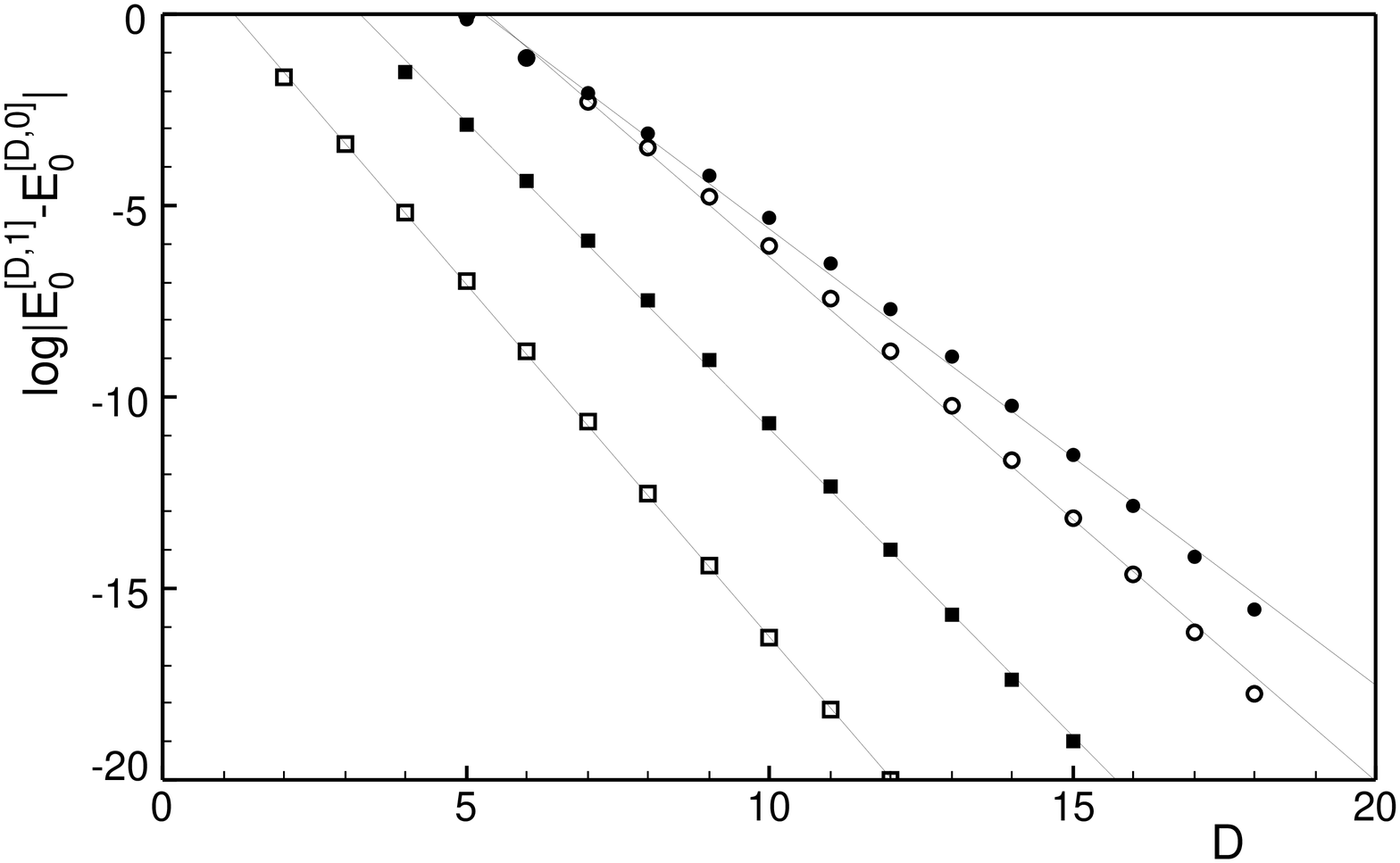}
\end{center}
\caption{Logarithmic convergence of the lower and upper bounds $%
\log|E_0^{[D,1]}-E_0^{[D,0]}|$ for the ground state of the double--well
oscillator (\ref{eq:Vx4x2}) for $\beta=-1$ (squares), $\beta=-5$ (filled
squares), $\beta=-10$ (circles), $\beta=-15$ (filled circles)}
\label{fig:DWLOG}
\end{figure}

Fig.~\ref{fig:DWLOG} shows $\log \left| E_{0}^{[D,1]}-E_{0}^{[D,0]}\right| $
for $D\leq 20$ and $\beta =-1,-5,-10,-15$, as well as straight lines
provided by straightforward linear regression of the data for each value of $%
\beta $. We appreciate that the rate of convergence decreases slightly and
the RPM sequences start at greater $D$ as $\beta $ becomes more negative.
However, in all those cases the convergence of the RPM is remarkable. We
obtain similar results for the first excited state. The rate of convergence
of the RPM bounds is considerably greater to that of the moment method\cite
{H92}. However, the latter method is more general and can be applied to a
wider variety of problems.

Table~\ref{tab:E0E1} shows the first two eigenvalues obtained from Hankel
determinants with $d=0$, $d=1$, and $D\leq 20$. The upper and lower bounds
agree to $20$ digits for all those values of $\beta $ except for $\beta =-15$%
. In this case we estimate $-50.8413872843819547<E_{0}<-50.8413872843819543$
and $-50.8413872841870053<E_{1}<-50.8413872841870051$.

It is worth mentioning that present approach is based on a Taylor expansion
of the solution about the top of the barrier, and it is well known that it
is more convenient to expand about one of the minima. However, in the latter
case we would have a nonsymmetric potential--energy function that we do not
discuss here.

\section{Conclusions\label{sec:conclusions}}

In this paper we have presented an approach for the accurate calculation of
eigenvalues and eigenfunctions of separable quantum--mechanical problems. In
this introductory communication we have restricted to one--dimensional
models with symmetric (even--parity) potential--energy functions. The
generalization to other cases is straightforward and will be discussed in
subsequent articles.

This presentation appears rather mathematical at first sight, but one must
keep in mind that the Schr\"{o}dinger equation for anharmonic oscillators
have many applications in the study of, for example, oscillatory phenomena%
\cite{CDL77}. Therefore, the RPM may be of practical utility in such
studies, particularly because its remarkable rate of convergence enables one
to obtain accurate results easily. We would not say that the RPM is the best
choice for all possible applications, but it is worth taking into
consideration that it does not require the calculation of matrix elements
which may be difficult for some potentials and only requires their Taylor
expansions. Besides, in some cases, like the anharmonic oscillator (\ref
{eq:V_x2_x4}), the RPM yields simple implicit equations that are reasonably
accurate for all values of the potential parameters.

But the main purpose of this paper is to discuss an approximate method that
does not share many features with those presented in standard textbooks on
quantum chemistry and quantum mechanics\cite{P68,CDL77}. It is those
peculiar features of the RPM that in our opinion could make it most
interesting for students with some knowledge in mathematics and physics. In
particular, the RPM is suitable for training students in the use of CAS that
facilitate most of the required algebra and analysis.

\begin{table}[H]
\caption{Ground--state eigenvalues of the models in equations (\ref
{eq:V_mod_PT}) and (\ref{eq:V_PT})}
\label{tab:spurious}
\begin{tabular}{ccc}
\hline
$D$ & $E_0(\ref{eq:V_mod_PT})$ & $-E_0(\ref{eq:V_PT})$ \\ \hline
2 & -3.9988835968549022343 & -8.9983715148790601222 \\
3 & -3.9999985313068943510 & -8.9999994018534834235 \\
4 & -3.9999999984549959290 & -8.9999999999381908008 \\
5 & -3.9999999999984803299 & -8.9999999999999974948 \\
6 & -3.9999999999999985472 & -9.0000000000000000000 \\
7 & -3.9999999999999999986 & -9.0000000000000000000 \\
8 & -4.0000000000000000000 & -9.0000000000000000000 \\ \hline
\end{tabular}
\end{table}

\begin{table}[H]
\caption{Ground-- and first--excited--state eigenvalues of the double--well
oscillator (\ref{eq:Vx4x2})}
\label{tab:E0E1}%
\begin{tabular}{|D{.}{.}{3}D{.}{.}{20}D{.}{.}{20}|}
\hline
  \multicolumn{1}{|c}{$\beta$} & \multicolumn{1}{c}{$E_0$} &
  \multicolumn{1}{c|}{$E_1$} \\ \hline
  -1 & 0.65765300518071512306 & 2.8345362021193042147 \\
  -5 & -3.4101427612398294753 & -3.2506753622892359802 \\
  -10 & -20.633576702947799150 & -20.633546884404911079 \\
  -15 & -50.8413872843819543 &   -50.8413872841870051 \\ \hline
\end{tabular}
\end{table}

\end{document}